\documentclass[12pt]{article}
\usepackage{wider}
\begin{document}

\title{\bf Locality  in the Everett Interpretation of 
Heisenberg-Picture Quantum Mechanics\thanks{
This work was sponsored by the Department of the Air Force under Contract 
F19628-00-C-0002.  Opinions, interpretations, conclusions, and 
recommendations are those of the author and are not necessarily endorsed 
by the United States Air Force.
}
}
\author{
Mark A. Rubin\\
\mbox{}\\   
Lincoln Laboratory\\ 
Massachusetts Institute of Technology\\  
244 Wood Street\\                         
Lexington, Massachusetts 02420-9185\\      
{\tt rubin@ll.mit.edu}\\ 
}
\date{\mbox{}}
\maketitle

\begin{abstract}

Bell's theorem depends crucially on counterfactual reasoning, and is
mistakenly interpreted as ruling out a local explanation for the
correlations which can be observed between the results of measurements
performed on spatially-separated quantum systems. But in fact the
 Everett interpretation of  
quantum mechanics, in the Heisenberg picture,  provides an alternative local explanation for such correlations.  Measurement-type interactions lead, not to many worlds
 but, rather, to many local copies of experimental systems and the observers who measure 
their properties.  Transformations of the 
Heisenberg-picture operators corresponding to the properties of  these systems and observers,
induced by measurement  interactions, ``label'' each copy and 
provide the mechanism which, e.g., ensures that each copy of one of the observers  
in an EPRB or  GHZM experiment will only 
interact with the ``correct'' copy of the other observer(s).  The conceptual problem of 
nonlocality is thus replaced with a conceptual problem of proliferating labels,
as correlated systems and observers
undergo measurement-type interactions with newly-encountered objects and instruments;
it is suggested that this problem may be resolved by considering quantum field theory 
rather than the quantum mechanics of particles.
 
\end{abstract}

\section{Introduction}

In the paper  in which he introduces what has come to be known as the Everett or
many-worlds interpretation of quantum mechanics, Everett (1957) 
states that ``fictitious 
paradoxes like that of Einstein, Podolsky, and Rosen which are concerned with \ldots 
correlated, noninteracting systems are easily investigated and clarified in the
present scheme.'' In the Everett interpretation the nonlocal notion of reduction
of the wavefunction is eliminated, suggesting that questions of the locality of
quantum mechanics might indeed be more easily addressed.     
On the
other hand, while wavefunctions do not suffer reduction in the Everett interpretation,
nonlocality nevertheless remains present in many accounts of this formulation. 
In DeWitt's (1970) 
often-quoted description, for example,  ``every quantum transition taking place on every star, in every galaxy, in every remote corner of the universe is splitting our local world on earth into  myriads of copies of itself.'' 
Contrary to this viewpoint, others  argue   
(Page, 1982; Tipler, 1986, 2000;  Albert and Loewer, 1988; Albert, 1992; Vaidman, 1994, 1998,
1999; Price, 1995; Lockwood, 1996; Deutsch, 1996; Deutsch and Hayden, 2000)   
that the Everett  interpretation can in fact {\em resolve}\/ the apparent contradiction between locality and quantum mechanics.   
In particular, Deutsch and Hayden (2000)  
apply the Everett interpretation to quantum mechanics in the Heisenberg picture, and show that in EPRB experiments,\footnote{
Deutsch and
Hayden (2000)  analyze a variant of the EPRB experiment in which, rather than passing
through rotated analyzer magnets, the correlated particles are themselves each independently rotated  before  their spins are measured. This setup yields the same correlations
as the usual one, but allows the flow of information to be tracked explicitly at each step.
}
  information regarding the  correlations between systems is encoded in the Heisenberg-picture operators corresponding to the observables of the systems, and is carried 
from system to system and from place to place in a  local manner. 
The picture which emerges is not one of measurement-type interactions ``splitting the
universe'' but, rather,  producing copies of the observers and observed physical systems 
which have interacted during the (local) measurement process  
(Tipler, 1986).

The purpose of the present paper is to summarize the formalism of measurement in the 
Everett interpretation of Heisenberg picture quantum mechanics and its application to the
EPRB and GHZM experiments,  to
emphasize the key aspects of this formulation   of quantum mechanics which allow it
to circumvent Bell's theorem 
(Bell, 1964) and to describe the conceptual framework---a 
``labeled copies interpretation''---which it seems to imply.
The information
carried in entangled Heisenberg-picture operators governs the nature of 
correlations observed between the states of entangled systems and the observers
who measure them. It is the existence of this mechanism for bringing about, in
a local manner, the perfect correlations which are observed, e.g., when the
analyzer magnets in the EPRB experiment are parallel, which vitiates the
reasoning which otherwise leads one to conclude that either Bell's theorem
must hold or nonlocal influences must come into play. 
Since, in this scenario, even an entity as simple as  an electron  carries
with it for eternity a record of other entities with which it has interacted,
this interpretation entails a conceptual difficulty of its own.  It is possible that
this difficulty may be less severe if quantized fields rather than particles are considered. 

In Section 2 below I  review the  aspects of Bell's theorem most relevant for the
EPRB and the GHZM experiments. In Section 3  the 
Everett model for quantum measurement is reviewed in the
original Schr\"{o}dinger picture formulation as well as in the  Heisenberg picture.
Section 4 contains an analysis of the EPRB and GHZM experiments from an Everett point
of view in the Heisenberg picture. In Section 5 I discuss the manner in which the
labeled copies interpretation of quantum mechanics avoids
Bell's theorem, the problem of label proliferation, and  the possible relevance
of quantum field theory for a solution. 

\section{Bell's Theorem and Counterfactual Reasoning}

There are many derivations of the many variants of Bell's theorem; here we
review one of the
simplest   
(Farris, 1995). 
Consider two observers performing Bohm's version  
(Bohm, 1951)  
of the Einstein-Podolsky-Rosen  
(Einstein et al., 1935) 
experiment (EPRB) on pairs of spin-1/2 particles in the singlet state, using pairs of Stern-Gerlach analyzer
magnets which can be independently oriented in one of three directions $120^{\circ}$\/
apart and perpendicular
to the particles' line of flight.  
Define the  quantity $Q$\/: 
\begin{equation}
Q=P_{uu}(0^\circ,120^\circ)+P_{uu}(120^\circ,240^\circ)+P_{uu}(240^\circ,0^\circ), \label{Qdef}
\end{equation}
where $P_{uu}(\phi_1,\phi_2)$\/ is the probability of both observers obtaining the result 
spin-up 
from a particle pair when  analyzers 1 and 2 are oriented in directions
$\phi_1$\/ and $\phi_2$\/ respectively.
  Each of the three probabilities on the right-hand side of
equation (1) can be determined experimentally to any desired degree of accuracy, by performing
many repetitions of the EPRB experiment  with the analyzers held in the appropriate
 directions.  There is no quantum-mechanical restriction on performing
these experiments because, in each case, we are measuring spin components
of two {\em different}\/ particles, so the measurements  commute.

However, whenever experiments are performed in which  both analyzer magnets have the same orientation $\phi$\/, we observe that
\begin{equation}
P_{uu}(\phi,\phi)=0
\end{equation}
for any and all choices of $\phi$\/.  That is, if the analyzer directions
are the same, we find that whenever a particle is deflected
in one direction by one of the analyzers, its partner is deflected in the opposite
direction  by the other analyzer.    
We find that the
correlations persist even when we consider only cases
in which the analyzer orientations come to be parallel by chance,
because they've been chosen at the 
last possible moment before the particles arrive by some random 
process (delayed-choice experiment). 
We are thus compelled to pose a ``bothersome question''   
(Mermin, 1990a):
What is the mechanism which brings about these correlations?
In answer, we adopt what seems the only explanation
open to us: Each particle, even before its spin  is measured by the analyzer,
carries with it information---``instruction sets,'' as termed by Mermin 
(1990a)---determining  what its response  will be to the analyzer   
in every possible orientation. 

Having
accepted this explanation, our doom is sealed.  For if this explanation holds, it
is a well-defined notion to talk about what {\em would}\/ have happened if an analyzer 
had been oriented other than as it actually was
in any given experiment.  
That is, we define
the quantity $P^{(1)}_{IS-ud}(\phi_1,\phi_2)$\/ to be the probability that particle 1
is carrying instructions to be deflected up by an analyzer with orientation $\phi_1$\/, 
and at the same time
is carrying instructions  to be deflected down by an  analyzer at orientation $\phi_2$\/. 
This cannot be measured directly; but, by the reasoning
above, it has the value
\begin{equation}
P^{(1)}_{IS-ud}(\phi_1,\phi_2) = P_{uu}(\phi_1,\phi_2).
\end{equation}
So, using this in (\ref{Qdef}),
\begin{equation}
Q=P^{(1)}_{IS-ud}(0^\circ,120^\circ) + P^{(1)}_{IS-ud}(120^\circ,240^\circ)+
            P^{(1)}_{IS-ud}(240^\circ,0^\circ),
\label{QIS}
\end{equation}
and since the probabilities which are being added on the right-hand side of (\ref{QIS}) are of mutually exclusive events (e.g.,   particle 1 is carrying instructions to be
deflected {\em either}\/ up 
{\em or}\/ down by a magnet with orientation $120^\circ$\/) we conclude
\begin{equation}
Q \leq 1. \label{Qcontrapred}
\end{equation}
This inequality contradicts the prediction obtained from a quantum-mechanical
calculation of $Q$\/ (see, e.g., Section \ref{EPRBe}  below), and it is the latter which
is borne out by actual experiments 
(Aspect et al., 1982; Weihs et al., 1998).

The arguments leading to Mermin's 
(1990b,c)  
three-particle version of the 
Green\-berger-Horne-Zeilinger  
(Greenberger et al., 1989,1990)  
experiment (GHZM)
 are similarly based on the need to explain perfect correlations. 
In this case, the fact that the results of spin  measurements made on one particle correlate perfectly with those made on two other particles drives us to conclude that the
results of spin measurements on all three particles are governed by instruction sets.
The three particles in question each have spin-1/2 and travel outward from a common 
source in three coplanar directions.  The spin of each particle is measured by an
analyzer magnet that can be oriented at any angle in the plane perpendicular to
the corresponding particle's line of flight.  (The $0^\circ$\/ direction is perpendicular
to the common plane of the particles' motion.) Quantum mechanics predicts (see Section
\ref{GHZMe}) that 
\begin{equation}
P_{eu}(0^\circ,90^\circ,90^\circ)=P_{eu}(90^\circ,0^\circ,90^\circ)
=P_{eu}(90^\circ,90^\circ,0^\circ)=0,
\end{equation}
where $P_{eu}(\phi_1,\phi_2,\phi_3,)$\/ is the probability that an even number of
spin measurements will be up.  Enumeration of the possible instruction sets that
could account for these results 
(Mermin, 1990b) 
leads  to the conclusion---here, not
an inequality, but an equality---that
\begin{equation}
P_{eu}(0^\circ,0^\circ,0^\circ)=0 \hspace*{5mm} \mbox{(instruction set prediction).} \label{GHZM_IS}
\end{equation}
However, a direct quantum-mechanical calculation of this quantity gives
a probability , not of zero, but of unity (see Section \ref{GHZMe}).

\section{Everett's Measurement Model}

\subsection{Schr\"{o}dinger Picture}

Everett's paper presents a model of  ideal
measurements  in quantum mechanics.  
Consider a  physical system ${\cal S}$\/ and an observer\footnote{The observer, of course,
is also a physical system!}  
 ${\cal O}$. The space of states of ${\cal S}$\/
is spanned by the eigenstates of a Hermitian operator $\widehat{a}$\/ with eigenvalues
$\alpha_i$\/:
\begin{equation}
\widehat{a}|{\cal S};\alpha_i\rangle=\alpha_i|{\cal S};\alpha_i\rangle,
\hspace*{5mm}i=1,\ldots,N.
\end{equation}
For simplicity we will assume that the eigenvalues $\alpha_i$\/ are nondegenerate.
Here and below, unless indicated otherwise, all operators are time-independent
Schr\"{o}dinger-picture operators.  

The space of states
of ${\cal O}$\/ is  spanned by the eigenstates of a Hermitian operator $\widehat{b}$\/
with eigenvalues $\beta_I$\/:
\begin{equation}
\widehat{b}|{\cal O};\beta_I\rangle=\beta_I|{\cal O};\beta_I\rangle,
\hspace*{5mm}I=0,\ldots,N.
\end{equation}
The eigenvalues $\beta_I$\/ correspond to distinct ``states of belief'' of the
observer ${\cal O}$\/ concerning the results of measurements made on the system
${\cal S}$\/ (${\cal O}$\/ could of course be a computer or other machine rather
than a conscious human), so we can take them to be nondegenerate, with
$\beta_0$\/ corresponding to the state of ignorance (no measurement yet made).

The interaction corresponding to the measurement of ${\cal S}$\/ by 
${\cal O}$\/  is
represented by a unitary time-evolution operator $\widehat{U}_M$\/ acting in the product
space of the state spaces of ${\cal S}$\/ and ${\cal O}$\/.  In order to
correspond to an ideal measurement (the only type considered in this paper),
$\widehat{U}_M$\/ must have the  property that if at time $t_1$\/
${\cal O}$ is in a state of ignorance and ${\cal S}$\/ is in a state where
the quantity represented by $\widehat{a}$\/ definitely has the value 
$\alpha_i$\/---i.e., 
\begin{equation}
|\psi;t_1\rangle=|{\cal O};\beta_0\rangle |{\cal S};\alpha_i\rangle,
\end{equation}
so
\begin{eqnarray}
\widehat{A}|\psi;t_1\rangle&=&\alpha_i|\psi;t_1\rangle,\\
\widehat{B}|\psi;t_1\rangle&=&\beta_0|\psi;t_1\rangle,
\end{eqnarray}
where
\begin{eqnarray}
\widehat{A}&\equiv&\widehat{a}\otimes\widehat{I}_{\cal S},  \label{A0}\\
\widehat{B}&\equiv&\widehat{I}_{\cal O}\otimes\widehat{b},  \label{B0}\\
\widehat{I}_{\cal S}&\equiv& \mbox{identity operator in space of states of ${\cal S}$}\/, 
     \label{I_S}\\
\widehat{I}_{\cal O}&\equiv& \mbox{identity operator in space of states of ${\cal O}$}
      \label{I_O}
\end{eqnarray}
---then the action of $\widehat{U}_M$\/ is given by 
\begin{equation}
|\psi;t_2\rangle  =  \widehat{U}|\psi;t_1\rangle
= |{\cal O};\beta_i\rangle |{\cal S};\alpha_i\rangle.
\end{equation}

Since $\widehat{U}_M$\/ is a linear operator, its effect
on a state in which ${\cal O}$\/ is ignorant and ${\cal S}$\/ is in an arbitrary
superposition of $\widehat{a}$\/ eigenstates,
\begin{equation}
|\psi;t_1\rangle=|{\cal O};\beta_0\rangle(\sum_i c_i|{\cal S};\alpha_i\rangle, \label{psi1}
\end{equation} 
is
\begin{equation}
|\psi;t_2\rangle=\widehat{U}_M|\psi;t_1\rangle
=\sum_i c_i |{\cal O};\beta_i\rangle |{\cal S};\alpha_i\rangle.
\end{equation}
The state $|\psi;t_2\rangle$\/ is said to be ``entangled,''  since it is not a product
of states in the respective state spaces of ${\cal S}$\/ and ${\cal O}$\/.

Therefore
\begin{equation}
\widehat{U}_M\/=\sum_i\widehat{u}_i \otimes \widehat{\Pi}_i,   \label{U_form}
\end{equation}
where $\widehat{\Pi}_i$\/ is the projection operator  into the $i^{th}$\/
$\widehat{a}$\/ eigenstate of  ${\cal S}$\/,
\begin{equation}
\widehat{\Pi}_i \equiv |{\cal S};\alpha_i\rangle \langle{\cal S};\alpha_i|,
\end{equation}
and $\widehat{u}_i$\/ are unitary operators in the space of states of ${\cal O}$\/
with the property
\begin{equation}
\widehat{u}_i|{\cal O};\beta_0\rangle=|{\cal O};\beta_i\rangle,
   \hspace*{5mm}i=1,\ldots,N.
\end{equation}
The action of $\widehat{u}_i$\/ on states $|{\cal O};\beta_I\rangle$\/, $I \neq 0$\/,
will not play a role in what follows. (In Section \ref{EPRB} below we  give a specific
example of operators $\widehat{u}_i$\/ for the case $N=2$\/.)

\subsection{Heisenberg Picture}

In the Heisenberg picture, time dependence is carried by the operators. Heisenberg-picture
operators will be distinguished by explicit time arguments.  At the initial time $t_0$\/
the Heisenberg-picture  operators are identical to their Schr\"{o}dinger-picture counterparts:
\begin{equation}
\begin{array}{rcccl}
\widehat{A}(t_0)&=&\widehat{A}&=&\widehat{I}_{\cal O} \otimes \widehat{a},\\
\widehat{B}(t_0)&=&\widehat{B}&=&\widehat{b} \otimes \widehat{I}_{\cal S}.
\end{array} \label{t0_ops}
\end{equation}
At time $t_2$\/, after ${\cal O}$\/ has measured ${\cal S}$\/, these operators
become, respectively,
\begin{equation}
\begin{array}{rcl}
\widehat{A}(t_2)&=&\widehat{U}_M^\dagger \, \widehat{A} \, \widehat{U}_M, \\
\widehat{B}(t_2)&=&\widehat{U}_M^\dagger \, \widehat{B} \, \widehat{U}_M.
\end{array} \label{t2_ops}
\end{equation}
Here $t_2 > t_1 >t_0$\/, and, for now,  it is assumed that no  interaction
takes place between $t_0$\/ and $t_1$\/.

From (\ref{A0}-\ref{I_O}), (\ref{U_form}), (\ref{t0_ops}), and (\ref{t2_ops}), 
\begin{eqnarray}
\widehat{A}(t_2)&=&\widehat{I}_{\cal O} \otimes \widehat{a},\\
\widehat{B}(t_2)&=&\sum_i \: \widehat{u}_i^\dagger \widehat{b} \widehat{u}_i \: \otimes \:
                     \widehat{\Pi}_i,
\end{eqnarray}
since
\begin{eqnarray}
\widehat{u}_i^\dagger  \widehat{u}_i &=& \widehat{I}_{\cal O},\\
\widehat{\Pi}_i^\dagger&=&\widehat{\Pi}_i,\\
\widehat{\Pi}_i \widehat{\Pi}_j &=& \widehat{\Pi}_i \delta_{ij},\\
\widehat{a}&=&\sum_i \alpha_i \widehat{\Pi}_i.
\end{eqnarray}

In the  Heisenberg picture operator the ${\cal S}$\/ observable 
$\widehat{A}(t)$\/ is the same after the measurement as before. 
(This will not be the case in general; see Section \ref{EPRBne}.)  However,  the
${\cal O}$\/ observable $\widehat{B}(t)$\/
 has become
entangled with ${\cal S}$\/, in that it is no longer in the form
(\ref{B0}), the tensor product of an operator acting on ${\cal O}$\/ states with the identity 
operator on ${\cal S}$\/ states, but instead acts nontrivially on the states of 
${\cal S}$\/.
This is the hallmark  of entanglement in the Heisenberg picture 
(d'Espagnat, 1995, Section 10.8). 
The Heisenberg picture state vector, on the other
hand, remains at all times equal to the nonentangled  
Schr\"{o}dinger picture time-$t_0$\/
state vector (\ref{psi1}).

\section{EPRB and GHZM Experiments}

In EPRB and GHZM experiments, 
the particles are prepared in a
state  in which they are entangled with each other before measurement.  A spin component of 
each of the particles is subsequently measured by a corresponding analyzer
magnet which can be at one of several orientations. As in the previous
section, the action of the unitary time evolution operator will first be determined
by working in the Schr\"{o}dinger picture and subsequently used to compute
the form of the time-dependent operators in the Heisenberg picture.   
For purposes
of computational convenience, all operator eigenstates employed will be time-independent
eigenstates of time-independent operators---i.e., Schr\"odinger-picture eigenstates.

\subsection{EPRB Experiment} \label{EPRB}
The two particles are denoted  ${\cal S}^{(p)}$\/
and the two observers  ${\cal O}^{(p)}$\/, $p=1,2.$\/
The space of states of ${\cal S}^{(p)}$\/ is spanned by eigenstates of the Hermitian
operator $\widehat{a}^{(p)}$\/. In this case $\widehat{a}^{(p)}$\/ is the $z$\/ component of the $p^{th}$\/-particle spin operator
\begin{equation}
\widehat{a}^{(p)}=\widehat{\sigma}_z^{(p)}, \hspace*{5mm} p=1,2, \label{adef}
\end{equation}
where spin is measured in units of $\hbar/2$.\/  The ${\cal S}^{(p)}$\/
eigenbasis thus given by
\begin{equation}
\widehat{a}^{(p)}|{\cal S}^{(p)}; \alpha_i\rangle=
          \alpha_i|{\cal S}^{(p)}; \alpha_i\rangle, \hspace*{5mm} i,p=1,2,
          \label{aaction}
\end{equation}
where
\begin{equation}
\alpha_1=+1, \hspace*{10mm}
\alpha_2=-1.\label{alphavalues}
\end{equation}
The space of states of ${\cal O}^{(p)}$\/  is spanned by 
\begin{equation}
\widehat{b}^{(p)}|{\cal O}^{(p)}; \beta_I\rangle=
          \beta_I|{\cal O}^{(p)}; \beta_I\rangle, \hspace*{5mm} I=0,1,2. \label{baction}
\end{equation}
The eigenvalue $\beta_0$\/ corresponds to the ignorant state of the observer. Eigenvalues
$\beta_1$\/ and $\beta_2$\/ correspond  to the observer ${\cal O}^{(p)}$\/
having measured the spin of particle ${\cal S}^{(p)}$\/ to be respectively
 up or down. 

Since  the EPRB experiment involves measurements in several  directions, we consider 
measurement interactions by the observers ${\cal O}^{(p)}$\/  using analyzer magnets
oriented along arbitrary
independent directions denoted by unit vectors $\vec{n}^{(p)}$\/,
\begin{equation}
\vec{n}^{(p)}=(n^{(p)}_x,n^{(p)}_y,n^{(p)}_z)=(\sin \theta^{(p)} \cos \phi^{(p)},
                             \sin \theta^{(p)} \sin \phi^{(p)},
                             \cos \theta^{(p)}). \label{unitvec}
\end{equation}  
The time-evolution operators
corresponding to these measurements are therefore
\begin{equation}
\begin{array}{rcl}
\widehat{U}^{(1)}_{M,\vec{n}^{(1)}}&=&\sum_i
    \widehat{u}^{(1)}_i \, \otimes \, \widehat{I}^{(2)}_{\cal O} \, \otimes \,
    \widehat{\Pi}^{(1)}_{i,\vec{n}^{(1)}}  \, \otimes \, \widehat{I}^{(2)}_{\cal S},\\
\widehat{U}^{(2)}_{M,\vec{n}^{(2)}}&=&\sum_i
    \widehat{I}^{(1)}_{\cal O}  \, \otimes  \, \widehat{u}^{(2)}_i \,  \otimes \,
     \widehat{I}^{(1)}_{\cal S} \, \otimes \, \widehat{\Pi}^{(2)}_{i,\vec{n}^{(2)}}, \label{UMs}
\end{array}
\end{equation}
where $\widehat{\Pi}^{(p)}_{i, \vec{n}^{(p)}}$\/  is the projection operator into the $i^{th}$\/
eigenstate of particle $p$\/ along direction $\vec{n}^{(p)}$\/,
\begin{equation}
\widehat{\Pi}^{(p)}_{i, \vec{n}^{(p)}}=|{\cal S}^{(p)},\vec{n}^{(p)};\alpha_i\rangle
                            \langle{\cal S}^{(p)},\vec{n}^{(p)};\alpha_i|. \label{projop}
\end{equation}
In terms of spin eigenstates defined with respect to the $z$\/ axis 
(Greenberger et al., 1990, Appendix A),
\begin{equation}
\begin{array}{rcl}
|{\cal S}^{(p)},\vec{n}^{(p)};\alpha_1\rangle&=&
\exp(-i\phi/2)\cos(\theta/2)|{\cal S}^{(p)};\alpha_1\rangle +
\exp(i\phi/2)\sin(\theta/2)|{\cal S}^{(p)};\alpha_2\rangle,\\
|{\cal S}^{(p)},\vec{n}^{(p)};\alpha_2\rangle&=&
-\exp(-i\phi/2)\sin(\theta/2)|{\cal S}^{(p)};\alpha_1\rangle +
\exp(i\phi/2)\cos(\theta/2)|{\cal S}^{(p)};\alpha_2\rangle,
\end{array} 
\end{equation}
so
\begin{equation}
\begin{array}{rcl}
\widehat{\Pi}^{(p)}_{1, \vec{n}^{(p)}}&=&
\cos^2(\theta^{(p)}/2) \, \widehat{\Pi}^{(p)}_{1} + 
\sin^2(\theta^{(p)}/2) \, \widehat{\Pi}^{(p)}_{2} \\
& &
+ \sin \theta^{(p)}\left(\exp (-i \phi^{(p)})  \,  \widehat{T}^{(p)}_{1-2} 
+ \exp(i \phi^{(p)})   \, \widehat{T}^{(p)}_{2-1}\right)/2,\\
\widehat{\Pi}^{(p)}_{2, \vec{n}^{(p)}}&=&
\sin^2(\theta^{(p)}/2) \, \widehat{\Pi}^{(p)}_{1} + 
\cos^2(\theta^{(p)}2) \, \widehat{\Pi}^{(p)}_{2} \\
& &
- \sin \theta^{(p)} \left(\exp (-i \phi^{(p)})  \,  \widehat{T}^{(p)}_{1-2} 
+ \exp(i \phi^{(p)})   \, \widehat{T}^{(p)}_{2-1}\right)/2,
\end{array} \label{projopdetail}
\end{equation} 
where
\begin{equation}
\begin{array}{rclcrcl}
\widehat{\Pi}^{(p)}_1&=&|{\cal S}^{(p)};\alpha_1\rangle \langle {\cal S}^{(p)};\alpha_1 |,&\;&
\widehat{\Pi}^{(p)}_2&=&|{\cal S}^{(p)};\alpha_2\rangle \langle {\cal S}^{(p)};\alpha_2 |,\\
\widehat{T}^{(p)}_{1-2}&=&|{\cal S}^{(p)};\alpha_1\rangle \langle {\cal S}^{(p)};\alpha_2 |,&\;&
\widehat{T}^{(p)}_{2-1}&=&|{\cal S}^{(p)};\alpha_2\rangle \langle {\cal S}^{(p)};\alpha_1 |.
\end{array} \label{moreprojopdetail}
\end{equation} 
The operators $\hat{u}^{(p)}_i$\/  have the properties
\begin{equation}
\begin{array}{lcr}
\hat{u}^{(p)}_1|{\cal O}^{(p)};\beta_I\rangle&=&|{\cal O}^{(p)};\beta_{I+1 \, \mbox{mod} \,3}\rangle,\\
\hat{u}^{(p)}_2|{\cal O}^{(p)};\beta_I\rangle&=&|{\cal O}^{(p)};\beta_{I-1 \, \mbox{mod} \,3}\rangle.
\end{array}\label{uaction}
\end{equation}
Of these properties, the relevant ones for what follows are those for $I=0$\/:
\begin{equation}
\begin{array}{lcr}
\hat{u}^{(p)}_1|{\cal O}^{(p)};\beta_0\rangle&=&|{\cal O}^{(p)};\beta_{1}\rangle,\\
\hat{u}^{(p)}_2|{\cal O}^{(p)};\beta_0\rangle&=&|{\cal O}^{(p)};\beta_{2}\rangle.
\end{array}\label{uaction0}
\end{equation}

In the Heisenberg picture, the average of the product of the results of the measurements made by
${\cal O}^{(1)}$\/   and ${\cal O}^{(2)}$\/ at time $t_2$\/ is
\begin{equation}
\langle B^{(1)}B^{(2)}\rangle(t_2) =
\langle \psi, t_0 |\widehat{B}^{(1)}(t_2)\widehat{B}^{(2)}(t_2)|\psi, t_0 \rangle, \label{Bprod}
\end{equation}
where $|\psi, t_0 \rangle$\/ is the state of ${\cal O}^{(1)}$\/,${\cal O}^{(2)}$\/,
${\cal S}^{(1)}$\/ and ${\cal S}^{(2)}$\/ at the initial time $t_0$\/,
and the $\widehat{B}^{(p)}(t_2)$\/ are the operators on states of ${\cal O}^{(p)}$\/
after both measurements have been made. (From (\ref{UMs}) we see that 
$\widehat{U}^{(1)}_{M,\vec{n}^{(1)}}$\/
and $\widehat{U}^{(2)}_{M,\vec{n}^{(2)}}$\/ commute, so
the order in
which the measurements are made is immaterial.) At time $t_0$\/,
\begin{equation}
\begin{array}{lcr}
\widehat{A}^{(1)}=\widehat{I}^{(1)}_{\cal O} \, \otimes \, \widehat{I}^{(2)}_{\cal O}  \, 		      \otimes \,\widehat{a}^{(1)} \, \otimes \,\widehat{I}^{(2)}_{\cal S},\\
\widehat{A}^{(2)}=\widehat{I}^{(1)}_{\cal O} \, \otimes \, \widehat{I}^{(2)}_{\cal O}  \, 			\otimes \, \widehat{I}^{(1)}_{\cal S} \, \otimes \,\widehat{a}^{(2)},\\
\widehat{B}^{(1)}=\widehat{b}^{(1)} \, \otimes \, \widehat{I}^{(2)}_{\cal O}  \, \otimes \, 
                   \widehat{I}^{(1)}_{\cal S} \, \otimes \,\widehat{I}^{(2)}_{\cal S},\\
\widehat{B}^{(2)}=\widehat{I}^{(1)}_{\cal O} \, \otimes \, \widehat{b}^{(2)}  \, \otimes \, 
                   \widehat{I}^{(1)}_{\cal S} \, \otimes \,\widehat{I}^{(2)}_{\cal S}.
\end{array}\label{ABt0}
\end{equation}

For the initial-time Heisenberg picture state, we will use the state
in which both observers are ignorant and each of the pair of particles has a well-defined spin, with particle 1 up with respect to the $z$\/
axis and particle 2 down:
\begin{equation}
|\psi,t_0\rangle=|{\cal O}^{(1)};\beta_0\rangle |{\cal O}^{(2)};\beta_0\rangle
	           |{\cal S}^{(1)};\alpha_1\rangle |{\cal S}^{(2)};\alpha_2\rangle.
      \label{psi0}
\end{equation}

\subsubsection{Nonentangled Particles} \label{EPRBne}

First consider the case in which the only interactions which take place
subsequent to time $t_0$\/ are  the measurements of the two particles by
the two observers.  The time evolution operator from time $t_0$\/ to time
$t_2$\/ is then
\begin{equation}
\widehat{U}
 \equiv \widehat{U}^{(2)}_{M,\vec{n}^{(2)}}\widehat{U}^{(1)}_{M,\vec{n}^{(1)}}. \label{UNE}
\end{equation}
The Heisenberg-picture observables at time $t_2$\/,
\begin{equation}
\begin{array}{rcccl} 
\widehat{A}^{(p)}(t_2) & = & \widehat{U}^{\dagger}\widehat{A}^{(p)}\widehat{U}
       & = &  \left(\widehat{U}^{(1)\dagger}_{M,\vec{n}^{(1)}}
              \widehat{U}^{(2)\dagger}_{M,\vec{n}^{(2)}}\right)\:
              \widehat{A}^{(p)} \:
              \left(\widehat{U}^{(2)}_{M,\vec{n}^{(2)}} 
              \widehat{U}^{(1)}_{M,\vec{n}^{(1)}}\right),  \hspace*{5mm}p=1,2,\\
\widehat{B}^{(p)}(t_2) & = & \widehat{U}^{\dagger}\widehat{B}^{(p)}\widehat{U}
       & = &  \left(\widehat{U}^{(1)\dagger}_{M,\vec{n}^{(1)}}
              \widehat{U}^{(2)\dagger}_{M,\vec{n}^{(2)}}\right)\:
              \widehat{B}^{(p)} \:
              \left(\widehat{U}^{(2)}_{M,\vec{n}^{(2)}}
              \widehat{U}^{(1)}_{M,\vec{n}^{(1)}}\right),  \hspace*{5mm}p=1,2,
\end{array}
\end{equation}
are therefore, using (\ref{UMs}),
\begin{eqnarray}
\widehat{A}^{(1)}(t_2) & = & \sum_{i,j} \;  
           \widehat{u}^{(1)\dagger}_i \widehat{u}^{(1)}_j  \;  \otimes  \;  
           \widehat{I}^{(2)}_{\cal O}  \;  \otimes  \; 
           \widehat{\Pi}^{(1)}_{i,\vec{n}^{(1)}} \widehat{a}^{(1)}
           \widehat{\Pi}^{(1)}_{j,\vec{n}^{(1)}}  \, \otimes \, 
           \widehat{I}^{(2)}_{\cal S},  \\
\widehat{A}^{(2)}(t_2) & = & \sum_{i,j} \;  
           \widehat{I}^{(1)}_{\cal O}  \;  \otimes  \; 
           \widehat{u}^{(2)\dagger}_i \widehat{u}^{(2)}_j  \;  \otimes  \; 
           \widehat{I}^{(1)}_{\cal S} \: \otimes \:
           \widehat{\Pi}^{(2)}_{i,\vec{n}^{(2)}} \widehat{a}^{(2)}
           \widehat{\Pi}^{(2)}_{j,\vec{n}^{(2)}}, \\
\widehat{B}^{(1)}(t_2) & = &
   \sum_{i} \; 
   \widehat{u}^{(1)\dagger}_i \widehat{b}^{(1)} \widehat{u}^{(1)}_i  \;  \otimes  \; 
   \widehat{I}^{(2)}_{\cal O}   \;  \otimes  \;  
   \widehat{\Pi}^{(1)}_{i,\vec{n}^{(1)}}   \;  \otimes  \;  \widehat{I}^{(2)}_{\cal S}, 
    \label{B1entangled}\\
\widehat{B}^{(2)}(t_2) & = &
   \sum_{i} \; 
   \widehat{I}^{(1)}_{\cal O}   \;  \otimes  \;  
   \widehat{u}^{(2)\dagger}_i \widehat{b}^{(2)} \widehat{u}^{(2)}_i  \;  \otimes  \; 
   \widehat{I}^{(1)}_{\cal S}  \;  \otimes  \; 
   \widehat{\Pi}^{(2)}_{i,\vec{n}^{(2)}}.   \label{B2entangled}
\end{eqnarray}

We see that in this case  ${\cal S}^{(p)}$\/ as well as  ${\cal O}^{(p)}$\
are  entangled.  From (\ref{aaction})-(\ref{baction}),
(\ref{projopdetail}), (\ref{moreprojopdetail}), (\ref{uaction0}), 
(\ref{Bprod}), (\ref{psi0}), (\ref{B1entangled}) and (\ref{B2entangled}),
the average value of the product of the spin measurements
is
\begin{equation}
\begin{array}{lrl}
 \langle B^{(1)}B^{(2)}\rangle (t_2) & & \\
 =  \langle \psi, t_0 |\sum_{i,j}  
\widehat{u}^{(1)\dagger}_i \widehat{b}^{(1)}\widehat{u}^{(1)}_i
\, \otimes \,
\widehat{u}^{(2)\dagger}_j \widehat{b}^{(2)}\widehat{u}^{(2)}_j
\, \otimes \,
 \widehat{\Pi}^{(1)}_{i,\vec{n}^{(1)}}  \, \otimes \,  
  \widehat{\Pi}^{(2)}_{j,\vec{n}^{(2)}}
|\psi, t_0 \rangle  & & \nonumber \\
 =  \sum_{i,j}\beta_i \beta_j 
\langle{\cal S}^{(1)};\alpha_1|\langle {\cal S}^{(2)};\alpha_2|
 \widehat{\Pi}^{(1)}_{i,\vec{n}^{(1)}}  \, \otimes \,  
  \widehat{\Pi}^{(2)}_{j,\vec{n}^{(2)}}
|{\cal S}^{(1)};\alpha_1\rangle |{\cal S}^{(2)};\alpha_2\rangle & &\nonumber \\
=
\left(\beta_1 \cos^2(\theta^{(1)}/2) + \beta_2 \sin^2(\theta^{(1)}/2)\right)
\left(\beta_1 \sin^2(\theta^{(2)}/2) + \beta_2 \cos^2(\theta^{(2)}/2)\right), & &
\end{array} \label{Bprodunentangled}
\end{equation}
while the individual expected spin measurements are 
\begin{equation} 
\begin{array}{rcl}
\langle B^{(1)}\rangle (t_2) &=&
\langle \psi, t_0 |\widehat{B}^{(1)}(t_2)|\psi, t_0 \rangle \nonumber\\
&=&
\langle \psi, t_0 |\sum_{i}  
   \widehat{u}^{(1)\dagger}_i \widehat{b}^{(1)} \widehat{u}^{(1)}_i \, \otimes \,
   \widehat{I}^{(2)}_{\cal O}  \, \otimes \, 
   \widehat{\Pi}^{(1)}_{i,\vec{n}^{(1)}}  \, \otimes \, \widehat{I}^{(2)}_{\cal S}
|\psi, t_0 \rangle\\
&=&\beta_1 \cos^2(\theta^{(1)}/2) + \beta_2 \sin^2(\theta^{(1)}/2),\\
\langle B^{(2)}\rangle (t_2) &=&
\langle \psi, t_0 |\widehat{B}^{(2)}(t_2)|\psi, t_0 \rangle \nonumber \\
&=&
\langle \psi, t_0 |\sum_{i}  
   \widehat{I}^{(1)}_{\cal O}  \, \otimes \, 
   \widehat{u}^{(2)\dagger}_i \widehat{b}^{(2)} \widehat{u}^{(2)}_i \, \otimes \,
   \widehat{I}^{(1)}_{\cal S} \, \otimes \,
   \widehat{\Pi}^{(2)}_{i,\vec{n}^{(2)}}
|\psi, t_0 \rangle \\
&=&\beta_1 \sin^2(\theta^{(2)}/2) + \beta_2 \cos^2(\theta^{(2)}/2).
\end{array}
\end{equation}
So,
\begin{equation} 
\langle B^{(1)}B^{(2)}\rangle (t_2)
=
\langle B^{(1)}\rangle (t_2)
\langle B^{(2)}\rangle (t_2),
\end{equation}
as of course should be the case for measurements of independent systems.

\subsubsection{Entangled Particles} \label{EPRBe}

Now suppose that between times $t_0$\/ and $t_1$\/  an interaction $\widehat{U}_E$\/
occurs between the two particles which takes the state 
$|{\cal S}^{(1)};\alpha_1\rangle |{\cal S}^{(2)};\alpha_2\rangle$\/ to the singlet state.
Specifically, let
\begin{equation}
\widehat{U}_E=\widehat{I}^{(1)}_{\cal O} \otimes \widehat{I}^{(2)}_{\cal O} \otimes
                   \widehat{u}_E ,   \label{U_E}
\end{equation}
where
\begin{eqnarray}
\widehat{u}_E |{\cal S}^{(1)};\alpha_1\rangle |{\cal S}^{(2)};\alpha_1\rangle
&=& |{\cal S}^{(1)};\alpha_1\rangle |{\cal S}^{(2)};\alpha_1\rangle, \\
\widehat{u}_E |{\cal S}^{(1)};\alpha_2\rangle |{\cal S}^{(2)};\alpha_1\rangle
&=&\left( 
|{\cal S}^{(1)};\alpha_1\rangle |{\cal S}^{(2)};\alpha_2\rangle + 
|{\cal S}^{(1)};\alpha_2\rangle |{\cal S}^{(2)};\alpha_1\rangle
\right)/\sqrt 2 ,   \\
\widehat{u}_E |{\cal S}^{(1)};\alpha_1\rangle |{\cal S}^{(2)};\alpha_2\rangle 
&=& \left( 
|{\cal S}^{(1)};\alpha_1\rangle |{\cal S}^{(2)};\alpha_2\rangle - 
|{\cal S}^{(1)};\alpha_2\rangle |{\cal S}^{(2)};\alpha_1\rangle
\right)/\sqrt 2 ,   \label{singlet}  \\
\widehat{u}_E |{\cal S}^{(1)};\alpha_2\rangle |{\cal S}^{(2)};\alpha_2\rangle
&=& |{\cal S}^{(1)};\alpha_2\rangle |{\cal S}^{(2)};\alpha_2\rangle .
\end{eqnarray}
The time evolution operator from time $t_0$\/ to time
$t_2$\/ is in this case
\begin{equation}
\widehat{U}^{\prime}
 \equiv  \widehat{U}^{(2)}_{M,\vec{n}^{(2)}}\widehat{U}^{(1)}_{M,\vec{n}^{(1)}}\widehat{U}_E,
\end{equation}
so the Heisenberg-picture observables at time $t_2$\/,
\begin{equation}
\begin{array}{rcccl}
\widehat{A}^{(p)\prime}(t_2) & = & \widehat{U}^{\prime\dagger}\widehat{A}^{(p)}\widehat{U}^{\prime}
       & = &  \left( \widehat{U}_E^{\dagger}
              \widehat{U}^{(1)\dagger}_{M,\vec{n}^{(1)}}
              \widehat{U}^{(2)\dagger}_{M,\vec{n}^{(2)}} \right)
              \: \widehat{A}^{(p)} \:
              \left( \widehat{U}^{(2)}_{M,\vec{n}^{(2)}}
              \widehat{U}^{(1)}_{M,\vec{n}^{(1)}}
              \widehat{U}_E\right),  \hspace*{5mm}p=1,2,\\
\widehat{B}^{(p)\prime}(t_2) & = & \widehat{U}^{\prime\dagger}\widehat{B}^{(p)}\widehat{U}^{\prime}
       & = &  \left( \widehat{U}_E^{\dagger}
              \widehat{U}^{(1)\dagger}_{M,\vec{n}^{(1)}}
              \widehat{U}^{(2)\dagger}_{M,\vec{n}^{(2)}}\right)
              \: \widehat{B}^{(p)} \:
              \left(\widehat{U}^{(2)}_{M,\vec{n}^{(2)}}
              \widehat{U}^{(1)}_{M,\vec{n}^{(1)}}
              \widehat{U}_E\right),  \hspace*{5mm}p=1,2,
\end{array}
\end{equation}
become 
\begin{eqnarray}
\widehat{A}^{(1)\prime}(t_2) & = & \sum_{i,j} \;
           \widehat{u}^{(1)\dagger}_i \widehat{u}^{(1)}_j \; \otimes \; 
           \widehat{I}^{(2)}_{\cal O} \; \otimes \;
           \widehat{u}^{\dagger}_E
           \left(
           \widehat{\Pi}^{(1)}_{i,\vec{n}^{(1)}} \widehat{a}^{(1)}
           \widehat{\Pi}^{(1)}_{j,\vec{n}^{(1)}}  \; \otimes \; 
           \widehat{I}^{(2)}_{\cal S}  
           \right)
           \widehat{u}_E,
           \\
\widehat{A}^{(2)\prime}(t_2) & = & \sum_{i,j} \;
           \widehat{I}^{(1)}_{\cal O} \, \otimes \,
           \widehat{u}^{(2)\dagger}_i \widehat{u}^{(2)}_j \; \otimes \;
           \widehat{u}^{\dagger}_E
           \left(
           \widehat{I}^{(1)}_{\cal S} \; \otimes \;
           \widehat{\Pi}^{(2)}_{i,\vec{n}^{(2)}} \widehat{a}^{(2)}
           \widehat{\Pi}^{(2)}_{j,\vec{n}^{(2)}}
           \right)
           \widehat{u}_E           
           , \\
\widehat{B}^{(1)\prime}(t_2) & = &
   \sum_{i} \;
   \widehat{u}^{(1)\dagger}_i \widehat{b}^{(1)} \widehat{u}^{(1)}_i \; \otimes \;
   \widehat{I}^{(2)}_{\cal O}  \; \otimes \; 
   \widehat{u}^{\dagger}_E
   \left(
   \widehat{\Pi}^{(1)}_{i,\vec{n}^{(1)}}  \; \otimes \; \widehat{I}^{(2)}_{\cal S}
   \right)
   \widehat{u}_E, 
    \label{B1entangled2}\\
\widehat{B}^{(2)\prime}(t_2) & = &
   \sum_{i} \;
   \widehat{I}^{(1)}_{\cal O}  \; \otimes \; 
   \widehat{u}^{(2)\dagger}_i \widehat{b}^{(2)} \widehat{u}^{(2)}_i \; \otimes \;
   \widehat{u}^{\dagger}_E
   \left(
   \widehat{I}^{(1)}_{\cal S} \; \otimes \;
   \widehat{\Pi}^{(2)}_{i,\vec{n}^{(2)}}   
   \right)
   \widehat{u}_E.
\label{B2entangled2}
\end{eqnarray}
Using 
(\ref{aaction})-(\ref{baction}),
(\ref{projopdetail})-(\ref{moreprojopdetail}), (\ref{uaction0}), 
(\ref{Bprod}),  (\ref{psi0}), 
(\ref{B1entangled2}) and (\ref{B2entangled2}),
\begin{equation}
\begin{array}{rcl}
 \multicolumn{2}{l}{\langle B^{(1)\prime}B^{(2)\prime}\rangle (t_2) }& \\
 &\multicolumn{2}{r}{=  \langle \psi, t_0 |\; \sum_{i,j}  
\widehat{u}^{(1)\dagger}_i \widehat{b}^{(1)}\widehat{u}^{(1)}_i
\, \otimes \,
\widehat{u}^{(2)\dagger}_j \widehat{b}^{(2)}\widehat{u}^{(2)}_j
\, \otimes \,
   \widehat{u}^{\dagger}_E
   \left(
 \widehat{\Pi}^{(1)}_{i,\vec{n}^{(1)}}  \, \otimes \,  
  \widehat{\Pi}^{(2)}_{j,\vec{n}^{(2)}}
   \right)
   \widehat{u}_E
|\psi, t_0 \rangle  } \nonumber \\
& \multicolumn{2}{l}{=\left(
\left(\beta_1+\beta_2\right)^2 - 
\left(\beta_1-\beta_2\right)^2\vec{n}^{(1)}\cdot\vec{n}^{(2)}\right)/4.}
\end{array} \label{Bprodentangled}
\end{equation}

If the eigenvalues labeling the observers' states of awareness are chosen
to be $\beta_i=\alpha_i$\/, (\ref{Bprodentangled}) takes the well-known form   
(Greenberger et al., 1990, Appendix B)
\begin{equation}
\langle B^{(1)\prime}B^{(2)\prime}\rangle (t_2)_{\beta_1=1,\: \beta_2=-1} 
 =-\vec{n}^{(1)}\cdot\vec{n}^{(2)}.
\end{equation}
If $\beta_1=1$\/ and $\beta_2=0$\/, eq.(\ref{Bprodentangled}) is equal to the probability
that both observers find the particles they measure to be deflected in the spin-up
direction:
\begin{equation}
\begin{array}{rcl}
\langle B^{(1)\prime}B^{(2)\prime}\rangle (t_2)_{\beta_1=1,\: \beta_2=0}
&=&P_{uu}(\vec{n}^{(1)},\vec{n}^{(2)})\\
&=&\left(1-\vec{n}^{(1)}\cdot\vec{n}^{(2)}\right)/4.
\end{array} \label{Bprodprob}
\end{equation}
If the angle between $\vec{n}^{(1)}$\/ and $\vec{n}^{(2)}$\/is  $120^\circ$\/,
(\ref{Bprodprob}) has the value $3/8$\/, so the quantum-mechanical prediction for
$Q$\/ in eq. (\ref{Qdef}) is
\begin{equation}
Q=9/8,
\end{equation}
contradicting the prediction (\ref{Qcontrapred}) that $Q \leq 1$.\/

\subsection{GHZM Experiment} \label{GHZMsection}

We now consider three particles ${\cal S}^{(p)}$\/ and their corresponding observers
${\cal O}^{(p)}$\/, $p=1,2,3$\/. In addition, we will explicitly introduce an additional observer ${\cal O}^{(0)}$\/ who ascertains the states of awareness of the three
observers ${\cal O}^{(1)}$\/, ${\cal O}^{(2)}$\/ and ${\cal O}^{(3)}$\/, after
they have performed their respective spin measurements. The space of states of 
${\cal O}^{(0)}$\/
is spanned by the eigenstates of the Hermitian operator $\widehat{G}$\/:
\begin{equation}
\widehat{G}=\widehat{g} \otimes \widehat{I}^{(1)}_{\cal O} \otimes \widehat{I}^{(2)}_{\cal  O}
\otimes \widehat{I}^{(3)}_{\cal  O} \otimes \widehat{I}^{(1)}_{\cal S} \otimes \widehat{I}^{(2)}_{\cal  S} \otimes \widehat{I}^{(3)}_{\cal S},
\end{equation}
where
\begin{equation}
\widehat{g}|{\cal O}^{(0)};\gamma_I \rangle=\gamma_I|{\cal O}^{(0)};\gamma_I \rangle,
\hspace*{5mm} I=0,1,2. \label{Gaction}
\end{equation} 
We want $\gamma_0$\/ to correspond to ignorance, while $\gamma_1$\/ and  $\gamma_2$\/
correspond respectively to ${\cal O}^{(0)}$\/ having determined that ${\cal O}^{(p)}$\/, $p=1,2,3$\/
have observed an odd or even number of spin-up results. The interaction corresponding
to the measurement of ${\cal O}^{(p)}$\/  by ${\cal O}^{(0)}$\/ is
therefore
\begin{equation}
\widehat{V}=\sum_i \: \widehat{v}_i \: \otimes  \widehat{P}_i \: \otimes \: \widehat{I}^{(1)}_{\cal S} \: \otimes 
\: \widehat{I}^{(2)}_{\cal S} \: \otimes \: \widehat{I}^{(3)}_{\cal S},  
\end{equation}
where
\begin{equation}
\widehat{v}_i|{\cal O}^{(0)};\gamma_0 \rangle=|{\cal O}^{(0)};\gamma_i \rangle,
\hspace*{5mm} i=1,2. \label{vaction}
\end{equation} 
and $\widehat{P}_i$\/, $i=1,2$\/ are  the projection operators into
the spaces of states in which the three observers ${\cal O}^{(p)}$\/ perceive,
respectively, an odd and an even number of spin-up results:
\begin{eqnarray}
\widehat{P}_1&=&
      \widehat{p}^{(1)}_1 \: \otimes \: \widehat{p}^{(2)}_2 \: \otimes \widehat{p}^{(3)}_2                      	\: + \:
      \widehat{p}^{(1)}_2 \: \otimes \: \widehat{p}^{(2)}_1 \: \otimes \widehat{p}^{(3)}_2
      \: + \:
      \widehat{p}^{(1)}_2 \: \otimes \: \widehat{p}^{(2)}_2 \: \otimes \widehat{p}^{(3)}_1
      \: + \: \nonumber \\
  & & \widehat{p}^{(1)}_1 \: \otimes \: \widehat{p}^{(2)}_1 \: \otimes \widehat{p}^{(3)}_1,
\\                
\widehat{P}_2&=&
      \widehat{p}^{(1)}_1 \: \otimes \: \widehat{p}^{(2)}_1 \: \otimes \widehat{p}^{(3)}_2                                  	\: + \:
      \widehat{p}^{(1)}_1 \: \otimes \: \widehat{p}^{(2)}_2 \: \otimes \widehat{p}^{(3)}_1
      \: + \:
      \widehat{p}^{(1)}_2 \: \otimes \: \widehat{p}^{(2)}_1 \: \otimes \widehat{p}^{(3)}_1
      \: + \:  \nonumber \\
  & & \widehat{p}^{(1)}_2 \: \otimes \: \widehat{p}^{(2)}_2 \: \otimes \widehat{p}^{(3)}_2,
\end{eqnarray} \label{P_projop}
where
\begin{equation}
\widehat{p}^{(p)}_i = |{\cal O}^{(p)};\beta_i\rangle \langle {\cal O}^{(p)};\beta_i|,
\hspace*{5mm}p=1,2,3,\hspace*{5mm}i=1,2. \label{p_projop}
\end{equation}
If the measurement by   ${\cal O}^{(0)}$\/ of ${\cal O}^{(p)}$\/ takes place between
times $t_2$\/ and $t_3>t_2$\/, 
\begin{eqnarray}
\widehat{G}(t_3)&=&\left(
\widehat{U}_E^\dagger 
\; \widehat{U}^{(1)\dagger}_M \; \widehat{U}^{(2)\dagger}_M\; \widehat{U}^{(3)\dagger}_M
\; \widehat{V}^\dagger \: 
\right)
\widehat{G} \:
\left(
\widehat{V} \; \widehat{U}^{(3)}_M \; \widehat{U}^{(2)}_M \;\widehat{U}^{(1)}_M \;
\widehat{U}_E
\right)\\
&=&\sum_{i,j,k,l} \:
\widehat{v}^{\dagger}_i \widehat{g}\widehat{v}_i 
\: \otimes \:
\widehat{u}^{(1)\dagger}_j \widehat{u}^{(2)\dagger}_k \widehat{u}^{(3)\dagger}_l
\widehat{P}_i \; 
\widehat{u}^{(3)}_l \widehat{u}^{(2)}_k \widehat{u}^{(1)}_j \: \otimes \: \nonumber \\
& & 
\widehat{u}^{\dagger}_E\left(
\widehat{\Pi}^{(1)}_{j,\vec{n}^{(1)}} \otimes \widehat{\Pi}^{(2)}_{k,\vec{n}^{(2)}} \otimes \widehat{\Pi}^{(3)}_{l,\vec{n}^{(3)}}
\right)\widehat{u}_E
\end{eqnarray}
where, as in the previous section, $\widehat{U}_E=\ldots \otimes \widehat{u}_E$\/
takes a  nonentangled state of ${\cal S}^{(p)}$\/ to an entangled state, and
$\widehat{u}^{(p)}_i$\/ takes ${\cal O}^{(p)}$\/ from a state of ignorance to a state
of awareness $\beta_i$\/.

We take the state at the initial time $t_0$\/ to be one in which all four observers
are ignorant and in which all three particles have spin up with respect to their
respective $z$\/-axes:
\begin{equation}
|\psi_G,t_0\rangle=|{\cal O}^{(0)};\gamma_0\rangle 
                   \left(\prod_{p=1}^3|{\cal O}^{(p)};\beta_0\rangle\right)
                   \left(\prod_{p=1}^3|{\cal S}^{(p)};\alpha_1\rangle\right).
\end{equation}
Here we take the positive $z$\/ axis for each particle to be in the direction of its
motion, and the positive $x$\/ axis perpendicular to the plane in which their
motion lies (same direction for all particles).

The expected value of ${\cal O}^{(0)}$\/'s awareness at time $t_3$\/ is 
\begin{equation}
\begin{array}{lll}
\multicolumn{2}{l}{\displaystyle
\langle\psi_G,t_0|\widehat{G}(t_3)|\psi_G,t_0\rangle =}&\\
& 
\displaystyle
\gamma_1 \sum_{\stackrel{\{j,k,l\}}{\mbox{\scriptsize odd \# 1's}}}
\left(\prod_{p=1}^3\langle{\cal S}^{(p)};\alpha_1|\right)
\widehat{u}^{\dagger}_E\left(
\widehat{\Pi}^{(1)}_{j,\vec{n}^{(1)}} \otimes \widehat{\Pi}^{(2)}_{k,\vec{n}^{(2)}} \otimes \widehat{\Pi}^{(3)}_{l,\vec{n}^{(3)}}
\right)\widehat{u}_E
\left(\prod_{p=1}^3|{\cal S}^{(p)};\alpha_1\rangle\right)  
&+
\\
& 
\displaystyle
\gamma_2 \sum_{\stackrel{\{j,k,l\}}{\mbox{\scriptsize even \# 1's}}}
\left(\prod_{p=1}^3\langle{\cal S}^{(p)};\alpha_1|\right)
\widehat{u}^{\dagger}_E\left(
\widehat{\Pi}^{(1)}_{j,\vec{n}^{(1)}} \otimes \widehat{\Pi}^{(2)}_{k,\vec{n}^{(2)}} \otimes \widehat{\Pi}^{(3)}_{l,\vec{n}^{(3)}}
\right)\widehat{u}_E
\left(\prod_{p=1}^3|{\cal S}^{(p)};\alpha_1\rangle\right).
&
\end{array} \label{Gexpt}
\end{equation}
If the eigenvalues $\gamma_1$\/ and $\gamma_2$\/ have the respective values 0 and 1,
then the operator $\widehat{G}$\/ measures the probability of ${\cal O}^{(0)}$\/
determining that the ${\cal O}^{(p)}$\/'s  observe an even number of spin-up particles
during one run of the GHZM experiment:
\begin{equation}
\begin{array}{lcr}
\multicolumn{2}{l}{
\displaystyle
P_{eu}(\vec{n}^{(1)},\vec{n}^{(2)},\vec{n}^{(2)})=
} & \\
&\multicolumn{2}{r}{\displaystyle
\sum_{\stackrel{\{j,k,l\}}{\mbox{\scriptsize even \# 1's}}}
\left(\prod_{p=1}^3\langle{\cal S}^{(p)};\alpha_1|\right)
\widehat{u}^{\dagger}_E\left(
\widehat{\Pi}^{(1)}_{j,\vec{n}^{(1)}} \otimes \widehat{\Pi}^{(2)}_{k,\vec{n}^{(2)}} \otimes \widehat{\Pi}^{(3)}_{l,\vec{n}^{(3)}}
\right)\widehat{u}_E
\left(\prod_{p=1}^3|{\cal S}^{(p)};\alpha_1\rangle\right).}
\end{array} \label{PeuG}
\end{equation}

\subsubsection{Nonentangled Particles}

We first consider the case in which an entangling interaction among the
particles is absent, i.e.,
\begin{equation}
\widehat{u}_E= I_{\cal S}^{(1)} \otimes I_{\cal S}^{(2)} \otimes I_{\cal S}^{(3)}.
\end{equation}
Then, using the above equation with (\ref{projopdetail}), (\ref{moreprojopdetail}),
and (\ref{PeuG})
\begin{equation}
\begin{array}{lll}
\multicolumn{2}{l}{\displaystyle P_{eu}(\vec{n}^{(1)},\vec{n}^{(2)},\vec{n}^{(2)})=}& \\
& \cos^2(\theta^{(1)}/2)\cos^2(\theta^{(2)}/2)\sin^2(\theta^{(3)}/2)  + 
\cos^2(\theta^{(1)}/2)\sin^2(\theta^{(2)}/2)\cos^2(\theta^{(3)}/2) & + \\
& \sin^2(\theta^{(1)}/2)\cos^2(\theta^{(2)}/2)\cos^2(\theta^{(3)}/2)  + 
\sin^2(\theta^{(1)}/2)\sin^2(\theta^{(2)}/2)\sin^2(\theta^{(3)}/2), & 
\end{array}
\end{equation}
independent of the $\phi^{(p)}$\/'s. So, for {\em any} choices of the analyzer-magnet
orientations $\vec{n}^{(p)}$\/ perpendicular to the particles' respective
directions of motion ($\theta^{(p)}=\pi/2$\/),
\begin{equation}
P_{eu}(\phi^{(1)},\phi^{(2)},\phi^{(3)})=1/2.
\end{equation}

\subsubsection{Entangled Particles} \label{GHZMe}

On the other hand, if $\widehat{U}_E$\/ is such as to take the initial ${\cal S}^{(p)}$\/ state
to the GHZM state, i.e.,
\begin{equation}
\displaystyle
\widehat{u}_E\left(\prod_{p=1}^3|{\cal S}^{(p)};\alpha_1\rangle\right)=
\left(\frac{1}{\sqrt 2}\right)
\left(\left(\prod_{p=1}^3|{\cal S}^{(p)};\alpha_1\rangle\right) -
      \left(\prod_{p=1}^3|{\cal S}^{(p)};\alpha_2\rangle\right)\right),
\end{equation}
the probability of ${\cal O}^{(0)}$\/ determining that an even number of spin-up
measurements are made is, using the above equation with(\ref{projopdetail}), (\ref{moreprojopdetail}),
and (\ref{PeuG}),
\begin{equation}
\displaystyle
P_{eu}(\vec{n}^{(1)},\vec{n}^{(2)},\vec{n}^{(2)})=
\left(1 + \cos(\phi^{(1)} + \phi^{(2)} + \phi^{(2)})
               \sin(\theta^{(1)})\sin(\theta^{(2)})\sin(\theta^{(3)})\right)/2
\end{equation}
or, for $\theta^{(p)}=\pi/2$\/,
\begin{equation}
\displaystyle
P_{eu}(\phi^{(1)},\phi^{(2)},\phi^{(3)})=
\left(1 + \cos(\phi^{(1)} + \phi^{(2)} + \phi^{(2)})\right)/2.
\end{equation}
So an even number of spin-up measurements will never be found if one of the
analyzers is oriented perpendicular to the plane of the particles' motion,
\begin{equation}
P_{eu}(0^\circ,90^\circ,90^\circ)=P_{eu}(90^\circ,0^\circ,90^\circ)
=P_{eu}(90^\circ,90^\circ,0^\circ)=0,
\end{equation}
but, contrary to the prediction (\ref{GHZM_IS}) from instruction-set reasoning, 
an even number of spin-up measurements will {\em always}\/ be found if all analyzers are
oriented in the same sense perpendicular to this plane:
\begin{equation}
P_{eu}(0^\circ,0^\circ,0^\circ)=1. 
\end{equation}

\section{Discussion}

In the Heisenberg-picture formalism, the reason for the difference between
the correlations of the observers' measurements in the nonentangled and entangled
cases   
is the presence  in the
operators $\widehat{B}^{(p)}$\/, $\widehat{G}$\/ of different factors  acting in subspaces of  states 
pertaining, not to the observers but, rather, to the particles with which the observers
have interacted by virtue of the measurements they've performed.  These factors are in effect  ``labels'' which become attached to
the observers ${\cal O}^{(p)}$\/ , ${\cal O}^{(0)}$\/
after they have undergone the local interactions  $\widehat{U}^{(p)}_{M,\vec{n}^{(1)}}$\/,
$\widehat{V}$\/.
In the case  that, prior to any measurements, the particles are subject to a local 
entangling interaction
$\widehat{U}_E$\/, each of the particles ${\cal S}^{(p)}$\/ is labeled with a factor 
acting in the space of states of the other particle(s) with which it has interacted,
so the label which becomes attached to ${\cal O}^{(p)}$\/ 
after measuring the corresponding particle ${\cal S}^{(p)}$\/  
involves factors corresponding to the particle which the other observer(s) measured.
In the end the  observers  compare their observations by means of another
local interaction (${\cal O}^{(1)}$\/  interrogates ${\cal O}^{(2)}$\/, ${\cal O}^{(0)}$\/
 interrogates  ${\cal O}^{(1)}$\/, ${\cal O}^{(2)}$\/, and ${\cal O}^{(3)}$\/),
which has the effect of computing  quantities such as (\ref{Bprod}) and (\ref{Gexpt}).

The conceptual picture which emerges 
is thus the following: Interactions between
entities label those entities. The labels consist of modifications to the
Heisenberg-picture operators corresponding to the properties of the entities.
Measurement-type interactions (\ref{U_form}) transform the operators for the
states of awareness of observers into sums  of operators, each corresponding
to a distinct state of awareness of the observer, and each 
labeled with factors corresponding to the system which the observer measured,
as well as to other systems with which {\em that}\/ system has previously interacted.
These labels control the subsequent results of measurement involving the labeled operators,
including in particular measurements of correlations between the states of awareness
of observers who have measured particles which have previously interacted with
one another.

Bell's theorem (\ref{Qcontrapred}), (\ref{GHZM_IS}) is avoided because the counterfactual reasoning
which leads to it is not required and cannot be justified. In answer to the question  
``What is the mechanism which brings about these correlations?'' there exists 
an answer other than the existence of instruction sets.  Namely: When one of the
observers performing, say, an  EPRB experiment with both analyzer magnets  oriented in the same 
direction
measures the spin of one of the paired particles, that observer splits into noninteracting
copies, each copy labeled with information corresponding to the states of the observed
particle as well as to the state of the other particle. When the two observers---or,
more precisely, the two pairs of 
observer-copies---exchange 
information about the results of their measurements, it is the attached
labels which ensure that the ``correct'' copies of each of the observers interact; e.g.,
 preventing two observer-copies who have both observed spin-up from communicating.

To be {\em completely}\/ precise, we should say ``two pairs of 
{\em sets}\/ of observer-copies.'' This is necessary because, of course,
the analyzer magnets need not be oriented in the same direction.
When the analyzer magnets do not have the same
orientation, there are  four possible outcomes which the observers
can experience, with (in general) unequal probabilities for the ``same-spin'' and ``different-spin''
cases. To avoid being led to the conclusion that our formalism erroneously implies
equal probabilities for all outcomes regardless of the magnet orientations  
(Ballentine, 1973; Graham, 1973),
we  proceed
along the lines of Deutsch's (1985)  
modification to the Everett
interpretation and   regard,  for example, the third and fourth
lines of  eq.  (\ref{ABt0}) as respectively representing  continuous infinities of identical observers ${\cal O}^{(1)}$\/ and  ${\cal O}^{(2)}$\/. 
The two terms in eq. (\ref{B1entangled2}) or eq. (\ref{B2entangled2}) then  represent continuous infinities of two different types of observers (``saw-up'' and ``saw-down''),
and  the four terms in the operator in the second member of eq. (\ref{Bprodentangled})
represent  continuous infinities of four different types of pairs of observers
(''saw-up/saw-up,'' ``saw-up/saw-down,'' etc.).  The relative number of each type, as well as the
specific nature of each type (states of awareness of the observers),   
is governed by the expectation value of the corresponding term in the initial
state $|\psi_0, t_0 \rangle$\/. 

So, the splitting of each observer  into copies at each measurement interaction is 
represented by the local dynamics of the operators describing their states of awareness
relative to what they were at the initial time $t_0$\/; in particular,
the  possibilities  for interaction of observers of entangled systems are determined
by the labels attached to the operators. 
Determination of the  number of each type of observer-copy produced at each splitting, 
as well 
as the specific state of awareness of each type of observer-copy, involves information 
about the initial conditions of the system, information which in the Heisenberg
picture is contained in the time $t_0$\/ state vector. 
(DeWitt (1998)  
emphasizes that quantum systems are 
``described jointly by the dynamical variables and the state-vector.'')
 Just as observers or other
entities may be regarded as receiving and carrying with them, in a  local manner,
the labels described above, they may also be envisioned as  carrying with them in 
a similarly local manner the
requisite  initial-condition
information.

Since one cannot argue for the existence of counterfactual instruction sets, the conditions of Bell's
theorem do not apply.  Had angles  other than those that
actually were used been chosen
for the analyzer magnets, copies of each observer carrying labels 
appropriate to those angles would have resulted.
 There are indeed ``instruction sets'' present; but
they determine, not the results of experiments which were not performed but, rather,
the possibilities for interaction and information exchange between the Everett copies of the
observers who have performed the experiments. 

Bohr's reply to EPR can also be reinterpreted in the present context.
Regarding correlations at a distance, Bohr (1935) states that ``of course there
is in a case like that just considered no question of a mechanical disturbance of the system
under investigation during the last critical stage of the measuring procedure.
But even at this stage there is essentially the question of  {\em an influence on the very conditions which define the possible
types of predictions regarding the future behavior of the system.''}\/ The Everett splitting
and labeling of each observer  constitutes just such an influence, determining the possible types of
interactions with  physical systems and observers which the observer can experience
in the  future without in any way producing a ``mechanical disturbance'' of  distant entities.

The Everett interpretation in the  Heisenberg picture thus  
removes nonlocality from the list of conceptual problems of 
quantum mechanics. The idea of viewing the tensor-product factors
in the Heisenberg-picture operators as in some sense ``literally real''
introduces, however, a conceptual problem of its own.\footnote{The fact that   
the precise details of representation of the Heisenberg-picture
operators
depend, e.g., on the choice of initial time $t_0$\/ 
(d' Espagnat, 1995, 
Section 10.8) 
should  
{\em not}  be  a problem in viewing them as ``real,''
any more than, e.g.,  the fact that the components of the electromagnetic stress
tensor depend on the choice of Lorentz frame.}   
Entanglement via the introduction of nontrivial ``label''  factors
is not limited to interactions between two or three particles; each
particle of matter is labeled, for eternity, by all the particles with
which it has ever interacted. {\em What is the physical mechanism
by means of which all of this information is stored?}\/ 

The issue of ``where the labels are stored'' may seem
less problematic in the context of the Everett interpretation of
Heisenberg-picture quantum field theory.  After all, in quantum field theory, 
operators  corresponding to each species of particle and evolving according to local differential equations already reside at
each point in spacetime. (In the EPRB and GHZM experiments the particles in question
are considered to be distinguishable and so may be treated, for purposes of analyzing the experiments, as quanta of different fields.
More complicated objects, such as observers and magnets, might be approximated as
excitations of effective composite fields, following, e.g., Zhou et al. (2000).)

Even in the event that such a program for a literal, indeed mechanistic picture of measurement in quantum field theory cannot be realized,  it remains the case that  Everett's model for measurement in the Heisenberg picture  provides a
quantum formalism which is explicitly local and in which the problem of Bell's theorem  does
not arise.

\section*{Acknowledgment} I would like to thank Allen J. Tino for countless
quantum conversations.

\section*{References}

\begin{description}
\item[\rm Albert,~D.~Z. (1992).]  {\em Quantum Mechanics and Experience}\/ (Harvard University
Press, Cambridge, MA).
\item[\rm Albert,~D. and Loewer,~B. (1988).]  ``Interpreting the many worlds interpretation,''
{\em Synthese}\/ {\bf 77} 16.
\item[\rm Aspect,~A., Dalibard,~J. and Roger,~G. (1982).]  ``Experimental tests of Bell's
inequalities using time-varying analyzers,'' {\em Phys. Rev. Lett.}\/ {\em 47} 1804-1807.  
\item[\rm Ballentine,~L.~E.  (1973).]  ``Can the statistical
postulate of quantum theory be derived?---a critique of
the many-universes interpretation,'' 
{\em Found. Phys.}\/ {\bf 3} 229-240.
\item[\rm Bell,~J.~S.  (1964).]  ``On the Einstein-Podolsky-Rosen paradox,''
{\em Physics}\/ {\bf 1} 195-200, reprinted in J.~S.~Bell, {\em Speakable
and Unspeakable in Quantum Mechanics}\/ (Cambridge University Press, Cambridge, 1987).
\item[\rm  Bohm,~D. (1951).] {\em Quantum Theory}\/ (Prentice-Hall, Englewood Cliffs, NJ).
\item[\rm Bohr,~N. (1935).]  ``Can quantum-mechanical description
of reality be considered complete?''  {\em Phys. Rev.}\/ {\bf 48} 696-702, quoted in
Ch. 16 of J.~S.Bell, {\em Speakable and Unspeakable in Quantum Mechanics}\/
(Cambridge University Press, Cambridge, 1987).
\item[\rm d'Espagnat,~B. (1995).]  {\em Veiled Reality: An Analysis of Present-Day
Quantum Mechanical Concepts}\/ (Addison-Wesley, Reading, MA).
\item[\rm Deutsch,~D. (1985).] ``Quantum theory as a universal physical
theory,'' {\em Int. J. Theor. Phys.}\/ {\bf 24} 1-41.
\item[\rm Deutsch,~D.  (1996).] ``Reply to Lockwood,'' 
{\em Brit. J. Phil. Sci.}\/ {\bf 47} 222-228.
\item[\rm Deutsch,~D. and Hayden,~P. (2000).]  ``Information flow in entangled
quantum systems,'' {\em Proc. R. Soc. Lond.}\/ A456, 1759-1774;
quant-ph/9906007.
\item[\rm DeWitt,~B.~S. (1970).]  ``Quantum mechanics and reality,'' {\em Physics Today}
{\bf 32} 155-165.
\item[\rm DeWitt, B.  (1998).]  ``The quantum mechanics of isolated systems,''
{\em Int. J. Mod. Phys.}\/ {\bf A13} 1881-1916.
\item[\rm Einstein,~A., Podolsky,~B. and Rosen,~N. (1935).] ``Can quantum-mechanical description
of reality be considered complete?'' {\em Phys. Rev.}\/ {\bf 47} 777-780.
\item[\rm Everett~III,~H.  (1957).]   `` `Relative state' formulation of quantum mechanics,''
 {\em Rev. Mod. Phys.}\/ {\bf 29}  454-462. Reprinted in B.~S.~DeWitt and
N.~Graham, eds., {\em The Many Worlds Interpretation of Quantum Mechanics}
(Princeton University Press, Princeton, NJ, 1973). 
\item[\rm Farris,~W.~G. (1995).]  ``Probability in quantum mechanics,'' appendix to
D.~Wick, {\em The Infamous Boundary: Seven Decades of Heresy in Quantum Physics}
(Springer-Verlag, New York). 
\item[\rm Graham,~N. (1973).]  ``The measurement of relative frequency,'' in
B.~S.~DeWitt and N.~Graham, eds., {\em The Many Worlds Interpretation of Quantum Mechanics}
(Princeton University Press, Princeton, NJ). 
\item[\rm Greenberger,~D.~M., Horne,~M., Shimony,~A. and Zeilinger,~A. (1990).]  ``Bell's theorem without
inequalities,'' {\em Am. J. Phys.}\/ {\bf 58} 1131-1143. 
\item[\rm Greenberger,~D.~M., Horne,~M. and Zeilinger,~A. (1989).]  ``Going beyond Bell's
theorem,'' in M. Kafatos, ed., {\em Bell's Theorem, Quantum Theory, and Conceptions of the 
Universe}\/  (Kluwer Academic,  Dordrecht, The Netherlands).
\item[\rm Lockwood,~M. (1996).]  `` `Many minds' interpretations of quantum mechanics,''
{\em Brit. J. Phil. Sci.}\/ {\bf 47} 159-188.  
\item[\rm Mermin,~N.~D. (1990a).]  ``Spooky actions at a distance: mysteries of the quantum theory,'' in {\em The Great Ideas Today}\/ (Encyclopedia Britannica Inc., 1988), reprinted in
N.~D.~Mermin, {\em Boojums All The Way Through: Communicating Science in a Prosaic Age}\/
(Cambridge University Press, Cambridge). 
\item[\rm Mermin,~N.~D. (1990b).]  ``Quantum mysteries revisited,'' {\em Am. J. Phys.}\/
{\bf 58} 731-734.
\item[\rm Mermin,~N.~D. (1990c).] ``What's wrong with these elements of reality?,''
{\em Physics Today}\/ June 1990, 9-11. 
\item[\rm Page,~D.~N. (1982).]  ``The Einstein-Podolsky-Rosen physical reality is completely
described by quantum mechanics,'' {\em Phys. Lett. }\/ {\bf 91A} 57-60. 
\item[\rm Price,~M.~C. (1995).] ``The Everett FAQ,'' http://www.hedweb.com/manworld.htm.
\item[\rm Tipler,~F.~J. (1986).]  ``The many-worlds interpretation of quantum mechanics in quantum cosmology,'' in R.~Penrose and C.~J.~Isham, {\em Quantum Concepts in Space and
Time}\/ (Oxford University Press, New York). 
\item[\rm Tipler,~F.~J. (2000).] ``Does quantum nonlocality exist? Bell's theorem and the many-worlds interpretation,'' quant-ph/00003146.
\item[\rm Vaidman,~L. (1994).] ``On the paradoxical aspects of new quantum experiments,''
in D.~Hull et al., eds., {\em PSA 1994: Proceedings of the 1994 Biennial Meeting of the Philosophy of Science Association}\/ {\bf 1}, pp. 211-217 (Philosophy of Science Association, East Lansing, MI, 1994).
\item[\rm Vaidman,~L. (1998).] ``On schizophrenic experiences of the neutron or why we should believe in the many-worlds interpretation of quantum theory,'' {\em Int. Stud. Phil. Sci.}\/
{\bf 12}, 245; quant-ph/9606006.
\item[\rm Vaidman,~L. (1999).] ``The many-worlds interpretation of quantum mechanics,''
http://\-www.\-tau.\-ac.\-il/\-~vaidman/\-mwi/\-mwstr.\-tml.
\item[\rm Weihs,~G., Jennewien,~T., Simon,~C., Weinfurter,~H and Zeilinger,~A. (1998).]
``Violation of Bell's inequality under strict Einstein locality conditions,''
{\em Phys. Rev. Lett.}\/ {\bf 81} 5039-5043.
\item[\rm Zhou,~D.~L., Yu,~S.~X. and Sun,~C.~P. (2000).] ``Idealization second quantization
of composite particles,'' quant-ph/0012080.
\end{description}

\end{document}